\documentclass{nature}

\usepackage{caption}
\usepackage{graphicx}
\usepackage[scaled]{helvet}
\usepackage{amsmath}
\usepackage{amssymb}
\usepackage{bm}
\usepackage{xfrac}
\newcommand*{\myfont}{\fontfamily{phv}\selectfont}
\usepackage{hyperref}
\usepackage{anyfontsize}
\usepackage{ragged2e}
\usepackage{microtype}

\usepackage{color}

\definecolor{CLBlue}{rgb}{0, .25, .8}
\definecolor{MyBlue}{rgb}{0, .24, .40}
\definecolor{MyTurquoise}{rgb}{0, .53, .49}
\definecolor{MyGreen}{rgb}{0, .35, 0}
\definecolor{MyOrange}{rgb}{.8, .46, 0}
\definecolor{MyRed}{rgb}{.57, .07, 0}
\definecolor{MyPurple}{rgb}{.46, .1, .46}

\title{Emergence of criticality in models of real neurons}

\author{David P.~Carcamo$^{1,2,3}$ \& Christopher W.~Lynn$^{1,2,3}$}

\begin{document}

%\linenumbers

\maketitle

\begin{affiliations}
\item Department of Physics, Yale University, New Haven, CT 06520, USA
\item Quantitative Biology Institute, Yale University, New Haven, CT 06520, USA
\item Wu Tsai Institute, Yale University, New Haven, CT 06520, USA
\end{affiliations}

%\newpage

\noindent {\large \myfont \textbf{Abstract}}
\vspace{-28pt}

\noindent\rule{\textwidth}{.5pt}

\noindent Critical systems sit near boundaries between qualitatively distinct behaviors. When inferring models of neural activity, this proximity to criticality is thought to require the precise tuning of parameters. Here, we show that as the number of neurons increases, criticality can emerge naturally without fine-tuning. When computing observable statistics from parameters (the forward problem), some small regions in parameter space map to large regions in statistics space. These special parameters are precisely those near criticality. Thus, when inferring parameters from experimental measurements (the inverse problem), models concentrate near critical points, and this concentration becomes stronger as the system grows. We illustrate this flow toward criticality across many large-scale recordings in the mouse brain. In the Curie-Weiss model of Ising spins, we find that all of the recordings collapse to a first-order phase transition, despite substantial differences in the underlying systems. Together, these results suggest a resolution to the tension between criticality and fine-tuning in models of neural activity.

\newpage

\noindent {\large \myfont \textbf{Main}}
\vspace{-28pt}

\noindent\rule{\textwidth}{.5pt}

Biological networks---from flocks of birds and schools of fish to clusters of bacteria and genetic interactions---often exhibit signatures of criticality.\cite{mora2011biological, Bialek-02, puy2024signatures, chen2012scale, nykter2008gene} In the brain, populations of neurons spanning different animals and experimental settings produce avalanches of cascading activity.\cite{beggs2003neuronal, petermann2009spontaneous, Beggs_Timme_2012, fontenele2019criticality} The correlations between neurons extend over spatiotemporal scales,\cite{segev2002long, ringach2009spontaneous, stringer2019spontaneous, macdowell2020low} follow heavy-tailed distributions,\cite{bedard2006does, buzsaki2014log, lynn2024heavy, lynn_exact_2023} and display scale-free structure.\cite{meshulam2019coarse} Models inferred from neural data consistently lie near phase transitions, where small changes in model parameters produce large changes in collective activity.\cite{meshulam2025statistical, tkacik_thermodynamics_2015, lynn_exact_2023, di2026extended, carcamo_statistical_2024, pachitariu2026critical} Moreover, these models appear to move closer to criticality as the number of neurons increases.\cite{meshulam2025statistical, tkacik_thermodynamics_2015, lynn_exact_2023, di2026extended} In simulations, this proximity to criticality provides functional advantages, from improved sensitivity to optimized information processing.\cite{Shew_Yang_Petermann_Roy_Plenz_2009, Shew_Yang_Yu_Roy_Plenz_2011, haldeman2005critical, kinouchi2006optimal}

However, in the space of possible models, critical points occupy a vanishingly small volume.\cite{Sethna-01, mora2011biological, tkacik_thermodynamics_2015} This suggests that criticality requires the precise calibration of parameters, forcing the system to delicately balance between distinct regimes.\cite{chialvo2010emergent} This fine-tuning stands in stark contrast to the incredible diversity of neural systems across species, regions, and cell types. Significant efforts to address this tension have produced mechanisms for self-organization as well as alternative explanations for signatures of criticality.\cite{chialvo2010emergent, levina2007dynamical, rubinov2011neurobiologically, van1998chaotic, lynn2024heavy, moretti2013griffiths, schwab2014zipfs, aitchison2016zipfs, morrell2021latent, humplik2017probabilistic} Yet the question remains: In models of real neurons, does criticality require fine-tuning?

To answer this question, we examine the geometry of the mapping between experimental measurements and models of neural activity.\cite{mastromatteo2011criticality, amari2000methods} Although critical models occupy small regions in the space of parameters, as the number of neurons grows, these models map to large regions in the space of observable statistics. Thus, as experiments record from larger populations, many different measurements from a range of systems collapse to similar models near criticality.

\noindent {\myfont \large Minimal mapping from experiments to models}

\begin{figure}[t!]
\centering
\includegraphics[width = \textwidth]{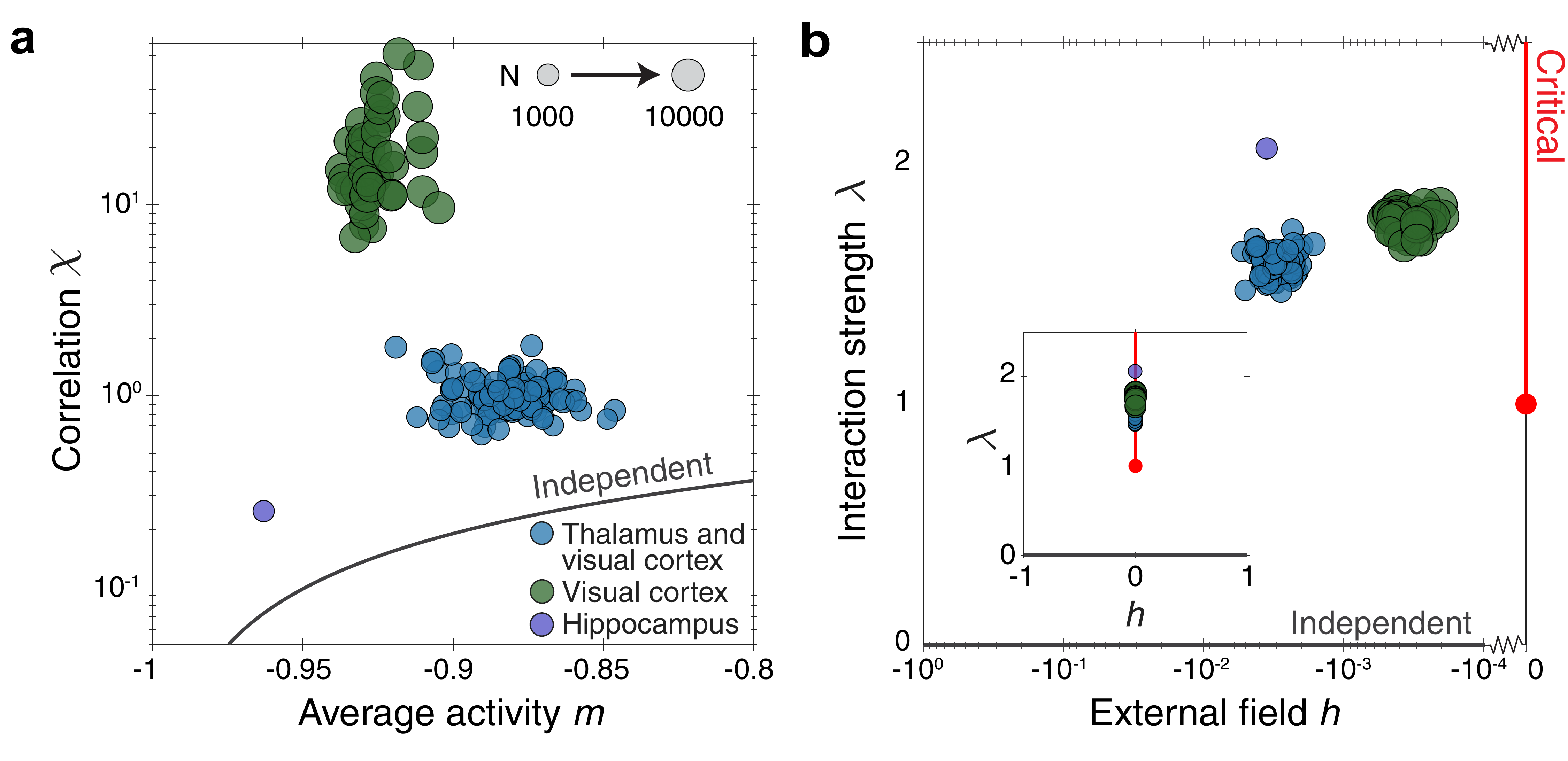} \\
\raggedright
\captionsetup{labelformat=empty}
{\spacing{1.25} \caption{\small \textbf{Fig.~\ref{fig_1} $|$ Neural activity maps to models near criticality.} \textbf{a}, Average activity $m$ and correlation strength $\chi$ for 146 recordings of neuronal activity: 100 electrophysiological recordings from thalamus and visual cortex (blue),\cite{allen_visual_coding_neuropixels, siegle2021survey} 45 calcium-imaging recordings from visual cortex (green),\cite{stringer_high-dimensional_2019} and one calcium-imaging recording from hippocampus (purple).\cite{Meshulam-03,gauthier2018dedicated} Marker size indicates the number of neurons $N$ in each population. Gray line represents statistics that map to independent models (with $\lambda=0$), defined by $\chi=1-m^2$. \textbf{b}, Inferred Ising parameters $h$ and $\lambda$ for each of the recordings in \textbf{a}. Red line defines the set of critical models in the thermodynamic limit $N \rightarrow \infty$, and gray line represents independent models ($\lambda=0$). Inset displays the same parameters with a linear $h$ axis, illustrating the collapse of the different recordings to critical models. \label{fig_1}}}
\end{figure}

To investigate this collapse in large populations of neurons, we focus on the simplest experimental measurements: the average activity and the strength of correlations. For a system of $N$ neurons ($i=1,\dots,N$), the state of the population at a given instant is defined by a binary vector $\bm{x}=\{x_i\}$, where each neuron is either active ($x_i=1$) or silent ($x_i=-1$). The instantaneous activity of the population is $\mu=\frac{1}{N}\sum_i x_i$, and experiments give us access to the mean activity $m=\frac{1}{N}\sum_i\langle x_i\rangle$, where $\langle \cdot \rangle$ denotes an experimental average. Similarly, the total strength of correlations is given by $\chi = \frac{1}{N}\sum_{i,j} \langle x_ix_j\rangle-\langle x_i\rangle\langle x_j\rangle = N\left(\langle\mu^2\rangle-m^2\right)$. The different combinations $(m,\chi)$ define our space of statistics, in which each point represents the possible outcome of an experiment. In Fig.~\ref{fig_1}a, we illustrate these statistics for 146 separate recordings from $N \approx 10^3$ to $10^4$ neurons in mice: 100 electrophysiological recordings from thalamus and visual cortex,\cite{allen_visual_coding_neuropixels, siegle2021survey} 45 calcium-imaging recordings from visual cortex,\cite{stringer_high-dimensional_2019} and one calcium-imaging recording from hippocampus.\cite{Meshulam-03, gauthier2018dedicated} These experiments span a large region in the space of statistics, with different activities $m$ and nearly two orders of magnitude in correlations $\chi$.

Every model of activity defines a distribution over collective states $P(\bm{x})$, but there are infinitely many models consistent with a given set of statistics $(m,\chi)$. To break this degeneracy, the only distribution justified by the measurements alone, without introducing any additional structure, is the one with maximum entropy,
\begin{equation}
P(\bm{x})=\frac{1}{Z}\exp\left[h\sum_i x_i+\frac{\lambda}{2N}\Big(\sum_i x_i\Big)^2\right],
\label{EQ:Model}
\end{equation}
where $Z$ is the normalizing partition function (Methods).\cite{thomas_m_cover_elements_2006, jaynes1957information} The parameters $h$ and $\lambda$ are Lagrange multipliers that must be computed so the model matches the observed statistics $m$ and $\chi$, respectively. This inference procedure defines a unique mapping from experimental measurements $(m,\chi)$ to model parameters $(h,\lambda)$.

Maximum entropy models such as Eq.~(\ref{EQ:Model}) have been applied across a wide range of neural systems to study the minimal consequences of experimental measurements.\cite{schneidman_weak_2006, meshulam2025statistical, tkacik_thermodynamics_2015, lynn_exact_2023, di2026extended, carcamo_statistical_2024} In the context of recurrent neural networks, $P(\bm{x})$ is the steady-state distribution for a system of logistic neurons with biases determined by $h$ and all-to-all interactions with strength determined by $\lambda$.\cite{hopfield1984neurons, kilian1996dynamic} In statistical physics, it is the Curie-Weiss model, which is a special case of the Ising model of interacting magnetic spins.\cite{ellis1978statistics, Sethna-01, di2026extended} Importantly, the critical behavior of the Curie-Weiss model is known exactly. In the thermodynamic limit $N\rightarrow \infty$, the system undergoes a second-order phase transition at the point $(h,\lambda) = (0,1)$. This is the classic Ising transition from disorder ($\lambda < 1$) to order ($\lambda > 1$) via spontaneous symmetry breaking.\cite{brush1967history, Sethna-01} Additionally, the system exhibits a line of critical points at $h = 0$ and $\lambda > 1$ that define a first-order phase transition.\cite{di2026extended} Crossing this line leads to a discontinuous jump from negative activity $m$ (for $h < 0$) to positive activity (for $h > 0$). Together, the second-order point and the first-order line represent the only critical models within the entire two-dimensional space of parameters.

Although we have defined a unique mapping between statistics $(m,\chi)$ and parameters $(h,\lambda)$, evaluating this map is computationally challenging. In maximum entropy models, calculations require averaging over all $2^N$ states, which typically restricts applications to populations of $N\sim 100$ neurons using Monte Carlo simulations.\cite{meshulam2025statistical} Here, given the simplicity of the Curie-Weiss model, we only need to sum over the $N+1$ discrete values of the population activity $\mu$, weighted by their multiplicities (Methods). This allows us to calculate exact statistics $(m,\chi)$ and numerically invert the model to infer exact parameters $(h,\lambda)$ for systems of up to $N \sim 10^4$ neurons.\cite{di2026extended}

Applying this method to each of the recordings, we arrive at 146 separate models (Fig.~\ref{fig_1}b). In each of these models, we find that $h < 0$, such that silence is favored over activity ($m < 0$), and $\lambda > 0$, such that neurons are positively correlated. Despite substantial differences in the measured statistics, brain regions, population sizes, experimental setups, and recording modalities, the inferred models collapse onto a narrow region in the space of parameters (Fig.~\ref{fig_1}b, \textit{inset}). Specifically, the different recordings are all described by similar models near criticality, with $h\approx 0$ and $\lambda \ge 1$. This collapse reinforces signatures of criticality that have been discovered in many distinct experiments and models of neural activity.\cite{beggs2003neuronal, petermann2009spontaneous, Beggs_Timme_2012, fontenele2019criticality, segev2002long, ringach2009spontaneous, stringer2019spontaneous, macdowell2020low, bedard2006does, buzsaki2014log, lynn2024heavy, lynn_exact_2023, meshulam2019coarse, meshulam2025statistical, tkacik_thermodynamics_2015, di2026extended, carcamo_statistical_2024, pachitariu2026critical} By contrast, if we shuffle the activity of each neuron to destroy correlations, then the statistics become similar while the models, which are no longer critical, remain distinct (Supplementary Information).

\noindent {\myfont \large Geometry of the forward map}

\begin{figure}[t!]
\centering
\includegraphics[width = \textwidth]{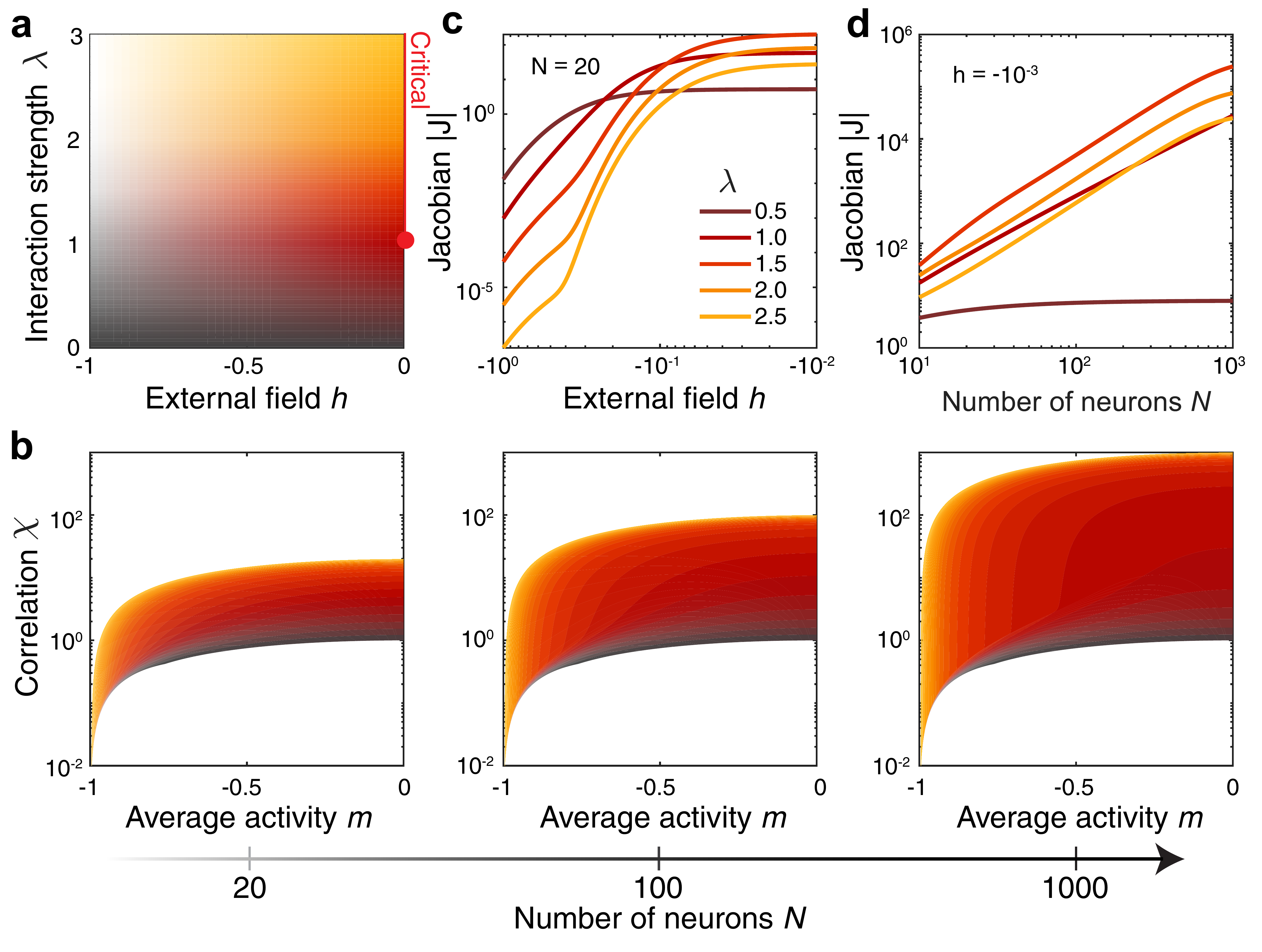} \\
\raggedright
\captionsetup{labelformat=empty}
{\spacing{1.25} \caption{\small \textbf{Fig.~\ref{fig_2} $|$ Expansion and contraction from parameters to statistics.} \textbf{a}, Region in the space of parameters with color indicating the interaction strength $\lambda$ and saturation indicating the external field $h$. Red line represents criticality in the thermodynamic limit. \textbf{b}, Regions in the space of statistics covered by the parameters in \textbf{a} for increasing numbers of neurons $N$. \textbf{c}, Scaling factor defined by the determinant of the Jacobian $|J|$ increases monotonically as a function of the external field $h$. Each line represents $N = 20$ neurons interacting with a different strength $\lambda$. \textbf{d}, Jacobian $|J|$ as a function of the system size $N$ at fixed $h=-10^{-3}$. For models near criticality ($\lambda \ge 1$), the Jacobian grows dramatically with the number of neurons.
\label{fig_2}}}
\end{figure}

To understand how disparate systems collapse to criticality, we must first understand the geometry of the forward map from model parameters to observable statistics. For a given number of neurons $N$, each set of parameters $(h,\lambda)$ defines a specific model [Eq.~(\ref{EQ:Model})], from which we can compute statistics $(m,\chi)$ (Methods). Consider the region of parameters outlined in Fig.~\ref{fig_2}a, which defines a manifold of models with different statistics. This mapping from parameters to statistics is highly nonlinear, with parameters near the critical line expanding to large volumes in the space of statistics and parameters far from criticality contracting to small volumes (Fig.~\ref{fig_2}b). This nonlinearity becomes even more dramatic as the number of neurons increases, with a large number of possible statistics captured only by a small number of critical models (Fig.~\ref{fig_2}b, \textit{right}).

%To understand this concentration, we first study the forward map from parameter space $(h,\lambda)$ to statistics space $(m,\chi)$. We color each point in parameter space according to its position in $h$--$\lambda$ space and map the full region into statistics space. The resulting image is strongly distorted, with regions near the critical line mapping to large regions of statistics space and regions farther from criticality mapping to much smaller regions. This distortion becomes stronger as the number of neurons increases (Figs.~\ref{fig_2}a--b).

To quantify the strength of this expansion and contraction, we evaluate the shift in statistics induced by small changes in parameters. For example, the fluctuation-dissipation relation reveals that the response in activity $m$ from a perturbation in $h$ is dictated by the strength of correlations, such that $\frac{\partial m}{\partial h} = \chi$. At the critical point $(h,\lambda) = (0,1)$, the correlation $\chi$ diverges as the number of neurons $N$ increases. Thus, for models near criticality, small changes in $h$ sweep over many different activities $m$, particularly as the population grows.

More generally, as parameters map to statistics, the scaling of volume is defined by the determinant of the Jacobian,
\begin{equation}
|J| = \left| \begin{array}{cc} \frac{\partial m}{\partial h} & \frac{\partial m}{\partial \lambda} \\ \frac{\partial \chi}{\partial h} & \frac{\partial \chi}{\partial \lambda} \end{array} \right|,
\label{EQ:J}
\end{equation}
which, for simplicity, we refer to as the Jacobian.\cite{mastromatteo2011criticality, amari2000methods} In the Curie-Weiss model, we can calculate $|J|$ exactly, even for large $N$ (Methods). As the external field $h$ increases, tending toward the critical line, the Jacobian grows monotonically (Fig.~\ref{fig_2}c). Far from criticality, volume contracts $(|J| < 1)$, while close to criticality, volume expands $(|J| > 1)$. Moreover, even if we hold parameters fixed near the critical line, as the number of neurons increases, the Jacobian grows by orders of magnitude (Fig.~\ref{fig_2}d). Thus, when mapping from parameters to statistics, not all models are created equal; a handful expand to cover many possible observations, and these are precisely the models near criticality.

\noindent {\myfont \large Growth induces flow toward criticality}

\begin{figure}
\centering
\includegraphics[width = \textwidth]{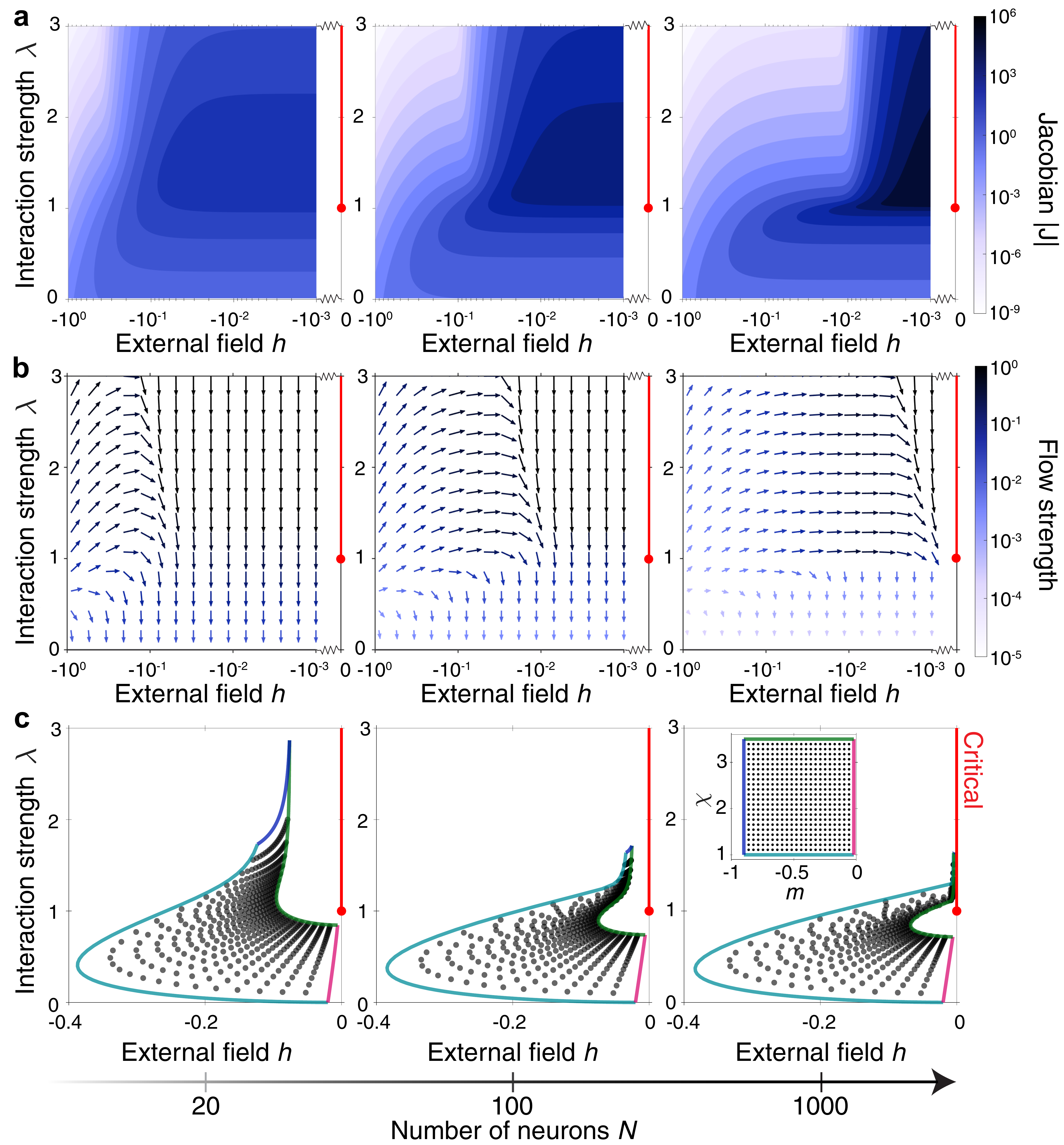} \\
\raggedright
\captionsetup{labelformat=empty}
{\spacing{1.25} \caption{\small \textbf{Fig.~\ref{fig_3} $|$ Large populations contract to critical models.} \textbf{a}, Jacobian $|J|$ as a function of the external field $h$ and interaction strength $\lambda$. As the number of neurons increases, the Jacobian concentrates near the critical line. \textbf{b}, Flow in parameters due to a fractional increase in system size $\frac{d(h,\lambda)}{d\log N}$ with statistics $(m,\chi)$ held fixed (Methods). Flow magnitude is indicated by vector color and length. As $N$ increases, many parameters flow toward the critical line. \textbf{c}, Inferred parameters corresponding to a fixed grid of statistics (\textit{inset}). For larger populations, many statistics contract to critical models. Columns indicate populations of increasing size $N=20$, $100$, and $1000$. \label{fig_3}}}
\end{figure}

When we say that criticality requires fine-tuning, we are implicitly placing a prior on the space of parameters. If this prior is uniform, then finding a model near criticality is surprising. However, if we define models based on their statistics, then this prior becomes highly non-uniform, as demonstrated in Fig.~\ref{fig_2}b. Instead, one should select a prior that is invariant under reparameterization. This is Jeffreys prior, which, for maximum entropy models, is proportional to the square root of the Jacobian $|J|$ (Methods).\cite{jeffreys1946invariant, amari2000methods} Even for small systems, the Jacobian is largest near the critical line, and this maximum sharpens as the number of neurons increases (Fig.~\ref{fig_3}a). From this perspective, models near criticality should become orders of magnitude more likely as populations grow.\cite{mastromatteo2011criticality}

We arrive at the same conclusion by examining the geometry of the inverse map from statistics to parameters. As $N$ increases, we have seen that parameters near criticality expand to cover many statistics (Fig.~\ref{fig_2}b). Conversely, as experiments grow, we should expect many distinct measurements to contract toward critical models, a phenomenon previously observed in financial data.\cite{mastromatteo2011criticality} We can investigate the contraction quantitatively by treating the number of neurons $N$ as a free parameter in the model [Eq.~(\ref{EQ:Model})]. Holding the statistics $(m,\chi)$ fixed, increasing $N$ induces a flow in the parameters $(h, \lambda)$, which we compute using the Hubbard-Stratonovich transformation (Methods). For interactions below the critical strength ($\lambda < 1$), we find that increasing $N$ leads to a decrease in $\lambda$, pushing models toward independence (Fig.~\ref{fig_3}b). By contrast, for strongly interacting systems ($\lambda \ge 1$), the external field $h$ increases as the population grows, pushing models toward the critical line. This flow to criticality can be studied using thermodynamic quantities such as the susceptibility and specific heat (Supplementary Information).\cite{tkacik_thermodynamics_2015} Here, we are able to investigate the flow directly by defining a grid of statistics $(m,\chi)$, which represents a collection of possible experiments (Fig.~\ref{fig_3}c, \textit{inset}). For small $N$, the inferred models span a relatively large region of parameters (Fig.~\ref{fig_3}c, \textit{left}). Yet as the number of neurons increases, even with the statistics held fixed, many of the inferred models become indistinguishable as they contract to criticality (Fig.~\ref{fig_3}c, \textit{right}).

\begin{figure}[t!]
\centering
\includegraphics[width = \textwidth]{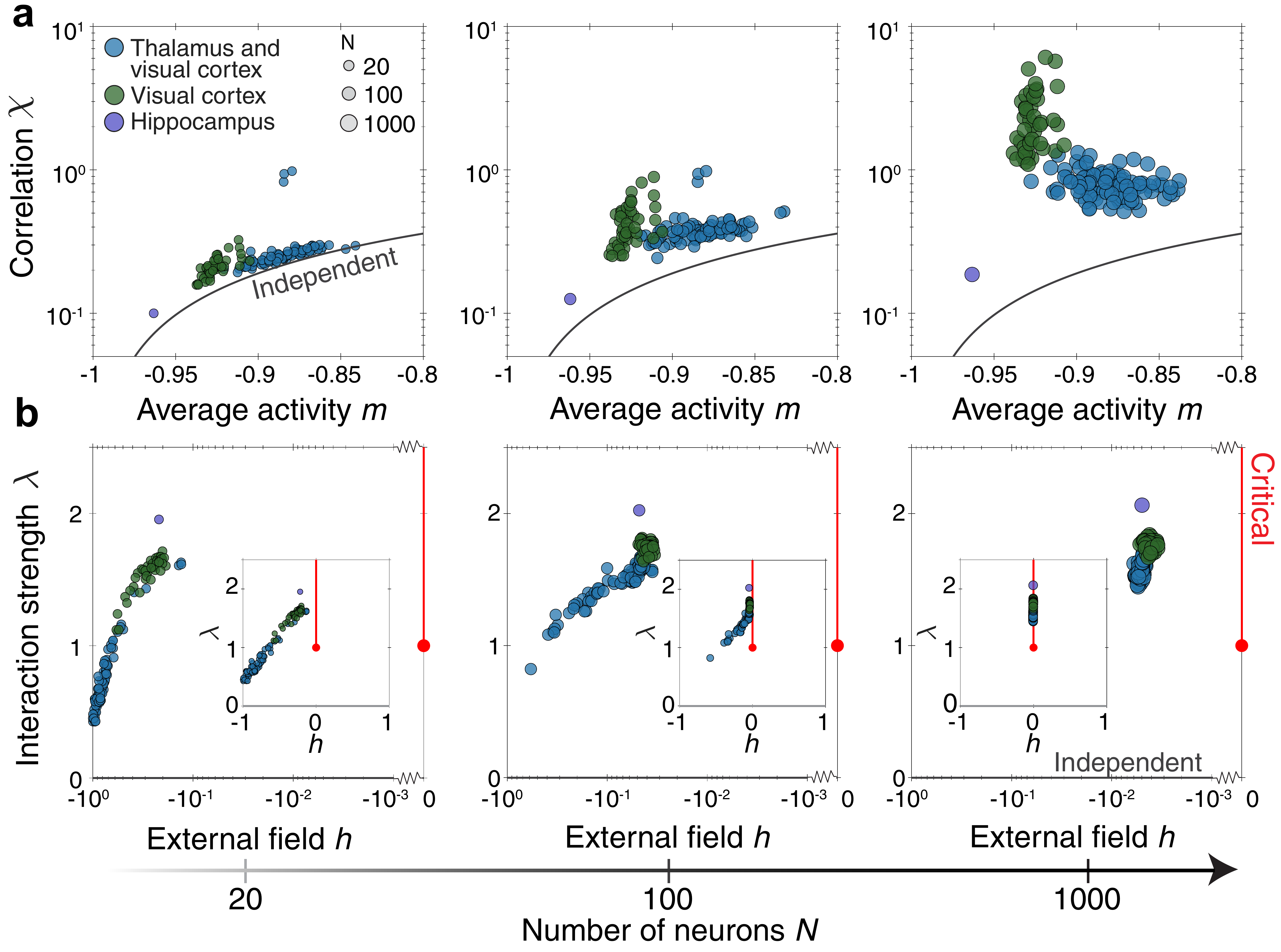} \\
\raggedright
\captionsetup{labelformat=empty}
{\spacing{1.25} \caption{\small \textbf{Fig.~\ref{fig_4} $|$ Models of neural activity flow toward criticality.} \textbf{a}, Average activity $m$ and correlation strength $\chi$ for subpopulations of increasing size $N$ sampled from 146 recordings of neural activity from the thalamus and visual cortex (blue),\cite{siegle2021survey, allen_visual_coding_neuropixels} visual cortex (green),\cite{stringer_high-dimensional_2019} and hippocampus (purple).\cite{Meshulam-03,gauthier2018dedicated} Marker size indicates the number of neurons $N$ in each subpopulation, and gray lines represent statistics that map to independent models. \textbf{b}, Inferred Ising parameters $h$ and $\lambda$ for each of the subpopulations in \textbf{a}. Red line defines the set of critical models in the thermodynamic limit $N \rightarrow \infty$, and gray line represents independent models ($\lambda=0$). Columns indicate populations of increasing size $N=20$, $100$, and $1000$. \label{fig_4}}}
\end{figure}

Ultimately, we seek to understand whether the same collapse occurs in neural activity. For each of the recordings in Fig.~\ref{fig_1}, we subsample the population at increasing sizes $N$. For each of these subpopulations, we measure the activity $m$ and correlation $\chi$, and then infer the corresponding parameters $(h,\lambda)$. For small $N$, correlations are weak, and the activity is nearly independent (Fig.~\ref{fig_4}a, \textit{left}). As the populations grow, correlations become stronger, and the recordings occupy an expanding volume of different statistics (Fig.~\ref{fig_4}a). In stark contrast, the corresponding models contract into a shrinking volume of similar parameters (Fig.~\ref{fig_4}b). Across all populations, the interactions rise above the critical strength $\lambda \ge 1$, and the external fields approach $h = 0$. Thus, despite expanding in the space of statistics, all of the recordings contract toward criticality in the space of models.

%We next ask whether the same behavior occurs in neural recordings. We subsample the populations in Fig.~\ref{fig_1} at increasing sizes and calculate $m$ and $\chi$ for each subsample. At small population sizes, the recordings lie close to the independent limit. As $N$ increases, the average activity remains roughly fixed within each recording, but the correlations become stronger. The recordings therefore remain distinct and occupy an increasingly broad region of statistics space (Fig.~\ref{fig_4}a).

%Despite the increasing separation of the neural statistics, the corresponding inferred models become more similar (Fig.~\ref{fig_4}b). The magnitude of the inferred external field decreases, and the parameters cluster progressively closer to the critical line across all brain regions and recording techniques. The models therefore do not approach criticality because the neural populations become more similar. Increasingly distinct statistics are represented by increasingly similar parameters. The geometry explains the direction of this flow, but it does not yet explain why the critical line is the limiting region as $N\to\infty$.

%\noindent {\myfont \large Thermodynamic limit}
\noindent {\myfont \large Thermodynamic collapse}

As the number of neurons increases, each individual interaction in the model becomes weaker, replaced by the average influence of a growing population. Thus, as experiments grow, one might expect a reduced mean-field theory to emerge.\cite{di2026extended} Here, we show that, while the simplest version of this idea fails, an improved theory captures the mapping between measurements and models. In doing so, we find that an entire two-dimensional region of statistics collapses to the one-dimensional critical line in the thermodynamic limit $N\rightarrow\infty$.

\begin{figure}
\centering
\includegraphics[width = \textwidth]{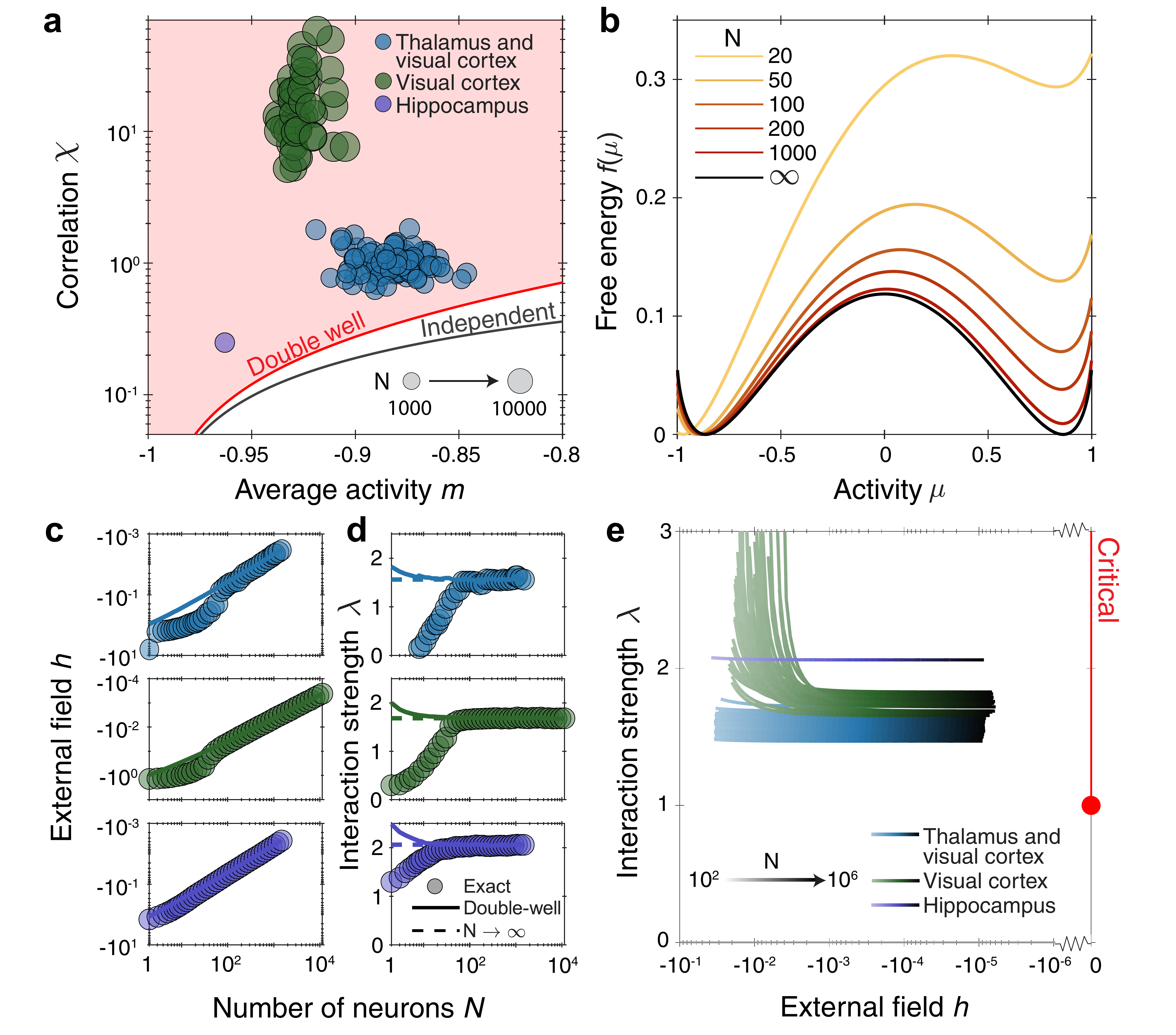} \\
\raggedright
\captionsetup{labelformat=empty}
{\spacing{1.25} \caption{\small \textbf{Fig.~\ref{fig_5} $|$ Collapse to criticality in the thermodynamic limit.} \textbf{a}, Average activity $m$ and correlation strength $\chi$ for the same recordings as Fig.~\ref{fig_1}. Red line defines the maximum correlations allowed under mean-field theory (Methods). Gray line defines statistics that map to independent models. Marker sizes indicate the number of neurons $N$ in each population. \textbf{b}, Free energy $f(\mu)$ as a function of the population activity for subpopulations of increasing size sampled from the hippocampal recording.\cite{Meshulam-03, gauthier2018dedicated} See Methods for free energy expression. \textbf{c-d}, External field (\textbf{c}) and interaction strength (\textbf{d}) inferred for subpopulations of increasing size sampled from the thalamus and visual cortex (\textit{top}),\cite{siegle2021survey, allen_visual_coding_neuropixels} visual cortex (\textit{middle}),\cite{stringer_high-dimensional_2019} and hippocampus (\textit{bottom}).\cite{Meshulam-03,gauthier2018dedicated} Markers represent exact parameters, solid lines represent the double-well approximation in Eq.~(\ref{EQ:double_well}), and dashed lines represent the thermodynamic limit of the double-well approximation. \textbf{e}, Double-well predictions for the external field and interaction strength [Eq.~(\ref{EQ:double_well})] corresponding to the statistics in \textbf{a} with increasing population size $N$. As $N\rightarrow \infty$, all recordings approach the critical line. \label{fig_5}}}
\end{figure}

Since the probability of each collective state $\bm{x}$ depends only on the population activity $\mu$, the model can be written $P(\bm{x}) \propto \exp[-Nf(\mu)]$, where $f(\mu)$ is the free energy per neuron (Methods). In the limit $N\rightarrow \infty$, this free energy has a unique minimum for any parameters $(h,\lambda)$ not on the critical line. In turn, this minimum dominates any average over $P(\bm{x})$, leading to the traditional mean-field theory,\cite{tanaka2000information} with the average activity defined self-consistently by $m = \tanh(\lambda m + h)$ and correlations by $\chi = \frac{1-m^2}{1 - \lambda(1-m^2)}$. As a direct consequence, for any activity $m$, mean-field theory sets a hard upper bound on the strength of correlations (Methods).\cite{di2026extended} However, when examining the neural recordings, we see that all of the measured correlations break this bound, some by multiple orders of magnitude (Fig.~\ref{fig_5}a). So where has the map between models and experiments broken down?

We have failed by focusing on models that are not critical. Consider the hippocampal recording.\cite{Meshulam-03, gauthier2018dedicated} For small subpopulations, the free energy maintains one global and one local minimum, but as $N$ increases, these minima become degenerate (Fig.~\ref{fig_5}b). This is the defining feature of a first-order phase transition. In order to map between parameters and statistics, we must therefore include both minima in any average over $P(\bm{x})$. The resulting double-well approximation predicts that, for measured statistics $(m,\chi)$, the corresponding parameters are given by
\begin{equation}
\label{EQ:double_well}
h = \frac{1}{Nm_0} \tanh^{-1}\left(\frac{m}{m_0}\right) \quad\quad \text{and} \quad\quad \lambda = \frac{1}{m_0} \tanh^{-1}(m_0),
\end{equation}
where $m_0 = \sqrt{m^2 + \chi/N}$ (Methods). We compare this prediction with the exact parameters inferred from different recordings (Fig.~\ref{fig_5}c,d). As populations grow, the double-well approximation becomes more accurate, eventually converging to the true parameters for $N \gtrsim 100$ neurons.

Upon inspection, the double-well approximation reveals something striking. For a given set of statistics $(m,\chi)$, the inferred interactions exceed the critical strength ($\lambda \ge 0$). Then, as the number of neurons increases, the external field vanishes ($h\rightarrow 0$), pushing models toward the critical line (Fig.~\ref{fig_5}e). This is precisely the flow that we observe in the neural recordings (Fig.~\ref{fig_4}b). Thus, in the thermodynamic limit $N \rightarrow \infty$, we find that a large region of possible statistics---which includes many different measurements of neural activity---collapses directly to the line of criticality.

\noindent {\myfont \large Discussion}

The brain exhibits several signatures associated with models poised near criticality.\cite{beggs2003neuronal, petermann2009spontaneous, Beggs_Timme_2012, fontenele2019criticality, segev2002long, ringach2009spontaneous, stringer2019spontaneous, macdowell2020low, bedard2006does, buzsaki2014log, lynn2024heavy, lynn_exact_2023, meshulam2019coarse, meshulam2025statistical, tkacik_thermodynamics_2015, di2026extended, carcamo_statistical_2024, pachitariu2026critical} How can such a diverse array of behaviors---across species, systems, and scales---correspond to such a narrow range of models? To answer this question, we study on the geometry of the map between experimental measurements and models of neural activity. Although critical models occupy a vanishingly small volume in the space of parameters, as the number of neurons grows, these models expand to cover a large volume of observable statistics.\cite{mastromatteo2011criticality} Thus, rather than requiring fine-tuning, criticality can emerge naturally in models of large-scale activity.

%We investigate more than one hundred recordings of large populations using minimal Curie-Weiss models of interacting neurons.\cite{allen_visual_coding_neuropixels, siegle2021survey, stringer_high-dimensional_2019, Meshulam-03,gauthier2018dedicated} Despite substantial differences in the underlying systems, all of the recordings collapse to similar models near a first-order phase transition. Together, these results suggest a resolution to the tension between criticality and fine-tuning in models of real neurons.

Focusing on the average activity $m$ and correlation strength $\chi$, the Curie-Weiss model defines the minimal mapping from experiments to models. Due to this simplicity, the critical behavior of the model is defined unambiguously, and we can exactly compute the mapping between statistics and parameters in very large populations (Methods). Although the Curie-Weiss model fails to capture many features of neural activity,\cite{di2026extended} we expect our central findings to hold much more broadly. For any statistics measured in experiments, the maximum entropy principle defines the most unbiased mapping to models.\cite{jaynes1957information, schneidman_weak_2006} In maximum entropy models, as the number of neurons increases, first-order phase transitions are defined by discontinuities in the statistics, and second-order phase transitions are defined by divergences in the Jacobian.\cite{Sethna-01, meshulam2025statistical, mastromatteo2011criticality} If we expect parameters to flow toward models where the Jacobian is large, then, quite generally, we should expect experimental measurements to contract toward criticality. Even in more complex descriptions of neural dynamics,\cite{pachitariu2026critical, chialvo2010emergent, van1996chaos} where criticality may not be defined, inferred models should concentrate where small changes in parameters produce large changes in behaviors. Investigating this contraction in different models and biological networks is an exciting direction.

Finally, we note that the brain can exhibit signatures of criticality without itself being critical. Indeed, some of these signatures can arise through correlated responses to latent variables or time-varying inputs.\cite{schwab2014zipfs, aitchison2016zipfs, morrell2021latent, humplik2017probabilistic} Even defining criticality for a system as complex as the brain remains an open challenge.\cite{Beggs_Timme_2012} However, when we fit models to neural data, our results suggest that criticality may not require fine-tuning. Instead, as experiments advance to record larger populations, many distinct measurements may collapse to similar models, thus providing simplified descriptions of neural systems.

\newpage

\noindent {\large \myfont \textbf{Methods}}
\vspace{-28pt}

\noindent\rule{\textwidth}{.5pt}

\begin{methods}

\subsection{Neural data.} We analyze neural activity from 146 electrophysiological and calcium-imaging recordings in previous experiments. The recordings were selected to be as large as possible, with no other selection criteria. The thalamus and visual cortex data were recorded electrophysiologically during a visual change-detection task as part of the Allen Institute Visual Behavior Neuropixels dataset.\cite{allen_visual_coding_neuropixels, siegle2021survey} We analyze 100 recordings, each containing $N= 1,705\pm355$ (mean $\pm$ SD) neurons from different brain regions. For each recording, spikes were sorted into $10$~ms time bins, which then defines the binary activity of each neuron. The visual cortex data were recorded using two-photon calcium imaging by Stringer \textit{et al}.\cite{stringer_high-dimensional_2019} The data consist of 45 recordings from seven awake mice, with $N=10,506\pm1,737$ neurons per recording (mean $\pm$ SD). Neural activity was measured at a sampling rate of approximately $1.5$ Hz while mice ran on an air-floating ball. Visual stimuli included natural images, distorted natural images, drifting gratings, and gray screens used to measure spontaneous activity. The hippocampal data were recorded using two-photon calcium imaging by Gauthier and Tank.\cite{gauthier2018dedicated}. The population consists of 1,485 neurons recorded at $30$~Hz from the CA1 region while the head-fixed mouse ran along a virtual track. In the calcium imaging data, fluorescence traces for each neuron were binarized into active or inactive states.

\subsection{Maximum entropy model.} Across all datasets, we represent the activity of neuron $i$ at time $t$ by a binary variable $x_i(t)\in \{-1,1\}$, where $x_i(t)=1$ corresponds to activity and $x_i(t)=-1$ to silence. The collective state of the population is then defined by a binary vector $\bm{x} = \{x_i\}$. For a recording of length $T$, we measure the average activity of each neuron $\langle x_i\rangle = \frac{1}{T}\sum_t x_i(t)$ and the correlation between each pair of neurons $\langle x_ix_j\rangle = \frac{1}{T}\sum_t x_i(t)x_j(t)$. From these measurements, we then compute the population-level average $m = \frac{1}{N}\sum_i \langle x_i\rangle$ and correlation $\chi = \frac{1}{N}\sum_{ij} \langle x_ix_j\rangle - \langle x_i\rangle\langle x_j\rangle$.

The only model $P(\bm{x})$ that is justified by these measurements, with no other sources of structure, is the one with maximum entropy.\cite{thomas_m_cover_elements_2006, jaynes1957information, schneidman_weak_2006} This distribution takes the form of the Curie-Weiss model in Eq.~(\ref{EQ:Model}), where
\begin{equation}
\label{EQ:partitionfunc}
Z=\sum_{\bm{x}}\exp\left[h\sum_i x_i+\frac{\lambda}{2N}\left(\sum_i x_i\right)^2\right]
\end{equation}
is the normalizing partition function. The parameters $h$ and $\lambda$ must be inferred so that the model matches the experimental activity $m$ and correlation $\chi$, respectively.

\subsection{Exact statistics and inference.} The probability of each state only depends on the instantaneous population activity $\mu = \frac{1}{N}\sum_i x_i$. When taking model averages, this allows us to replace the intractable sum over states $\bm{x}$ with a tractable sum over the $N+1$ distinct values of $\mu$, weighted by the number of states for each $\mu$. The partition function is given by
\begin{equation}
\label{EQ:Z}
Z = \sum_\mu C\Big(N,\frac{1+\mu}{2}N\Big) \exp\Big(Nh\mu+N\frac{\lambda}{2}\mu^2\Big),
\end{equation}
where $C(\cdot,\cdot)$ is the binomial coefficient. Similarly, the average activity and correlation strength can be efficiently computed using
\begin{align}
m &= \frac{1}{Z}\sum_\mu C\Big(N,\frac{1+\mu}{2}N\Big) \mu\exp\Big(Nh\mu+N\frac{\lambda}{2}\mu^2\Big),\\
\chi &= \frac{N}{Z}\sum_\mu C\Big(N,\frac{1+\mu}{2}N\Big) \mu^2\exp\Big(Nh\mu+N\frac{\lambda}{2}\mu^2\Big)-Nm^2.
\label{EQ:chi}
\end{align}
Eqs.~(\ref{EQ:Z})-(\ref{EQ:chi}) provide an exact solution for the forward map from parameters to statistics in very large populations.

To solve the inverse map from statistics to parameters, we note that at fixed $m$, there is a one-dimensional manifold of models along which $\chi$ varies, with parameters $h(\lambda)$ and $\lambda$.\cite{di2026extended} Parameter changes along this manifold satisfy
\begin{equation}
dm=\frac{\partial m}{\partial\lambda}d\lambda+\frac{\partial m}{\partial h}dh=0,
\end{equation}
so that
\begin{equation}
\frac{dh(\lambda)}{d\lambda}=-\left(\frac{\partial m}{\partial h}\right)^{-1}\frac{\partial m}{\partial\lambda} = -\frac{1}{\chi} \frac{\partial m}{\partial\lambda}.
\end{equation}
We numerically integrate this equation starting from the independent model $h(0)=\tanh^{-1}(m)$ until the model reaches the target value of $\chi$. In combination with Eqs.~(\ref{EQ:Z})-(\ref{EQ:chi}), this provides an exact solution to the inverse map.

\subsection{Jacobian, Jeffreys prior, and Fisher information.} The Jacobian in Eq.~(\ref{EQ:J}) is defined by the derivatives,
\begin{align}
\frac{\partial m}{\partial h} &= N (\langle \mu^2 \rangle - m^2) = \chi, \\
\frac{\partial m}{\partial \lambda} &= \frac{N}{2} (\langle \mu^3 \rangle - m \langle \mu^2 \rangle ), \\
\frac{\partial \chi}{\partial h} &= N^2 (\langle \mu^3 \rangle - 3 m \langle \mu^2 \rangle + 2 m^3), \\
\frac{\partial \chi}{\partial \lambda} &= \frac{N^2}{2} (\langle \mu^4 \rangle - \langle \mu^2 \rangle^2 -2 m \langle \mu^3 \rangle + 2 \langle \mu^2 \rangle m^2).
\end{align}
Note that each of the above model averages can be computed efficiently using the method described previously. Jeffreys prior defines the distribution over models that is invariant under changes in parameters $p(h,\lambda) \propto \sqrt{| I(h,\lambda)|}$, where $I(h,\lambda)$ is the Fisher information matrix.\cite{jeffreys1946invariant, amari2000methods} For the Curie-Weiss model, the Fisher information is given by\cite{mastromatteo2011criticality}
\begin{equation}
I = N^2 \begin{pmatrix} \langle \mu^2 \rangle - \langle \mu \rangle^2 & \frac{1}{2}\left(\langle \mu^3\rangle - \langle \mu^2\rangle\langle \mu \rangle\right) \\ \frac{1}{2}\left( \langle \mu^3\rangle - \langle \mu^2\rangle\langle \mu \rangle \right) & \frac{1}{4} \left( \langle \mu^4\rangle - \langle \mu^2 \rangle^2 \right) \end{pmatrix}.
\end{equation}
Note that the determinant of the Fisher information is proportional to that of the Jacobian, $|I| = \frac{N}{2} |J|$. Thus, Jeffreys prior is proportional to the square root of the Jacobian.

\subsection{Hubbard-Stratonovich transformation.} In order to treat the number of neurons $N$ as a free parameter in the model, we use the Hubbard-Stratonovich transformation.\cite{hubbard1959calculation} Replacing interactions in the model with an integral over a fluctuating field, we can rewrite the partition function exactly as,
\begin{equation}
Z = \sqrt{\frac{N}{2\pi \lambda }} 2^N \int d\psi e^{-Nl(\psi)},
\end{equation}
where
\begin{equation}
    l(\psi) = \frac{\psi^2}{2 \lambda } - \ln\cosh(h+\psi). 
\end{equation}
For any choice of $h$, $\lambda$, and $N$, the statistics can be computed numerically by evaluating the integrals,
\begin{equation}
    m = \frac{1}{\int d\psi e^{-Nl(\psi)}} \int d\psi e^{-Nl(\psi)} \tanh(h+\psi ),
\end{equation}
\begin{equation}
    \chi =  1-N m^2 + (N-1) \frac{1}{\int d\psi e^{-Nl(\psi)}} \int \tanh^2(h+\psi) e^{-Nl(\psi)}.
    \label{EQ:chiexact}
\end{equation}

\subsection{Parameter flow with increasing system size.} The system size now appears as a continuous variable, allowing us to differentiate the statistics with respect to $N$. Letting $\langle \cdot \rangle_\psi$ denote an average over the fluctuating field $P(\psi) \propto e^{-Nl(\psi)}$, we have
\begin{equation}
    \frac{\partial m}{\partial N}= m \langle l(\psi) \rangle_\psi - \langle \tanh(h+\psi) l(\psi)\rangle_\psi,
\end{equation}
\begin{equation}
    \frac{\partial \chi}{\partial N}= \langle \tanh(h+\psi)^2 \rangle_\psi + (N-1) \big(\langle \tanh(h+\psi)^2 \rangle_\psi \langle l(\psi) \rangle_\psi - \langle \tanh(h+\psi)^2 l(\psi)\rangle_\psi\big) - m^2 -2Nm\frac{\partial m}{\partial N}.
\end{equation}
To understand how models flow when the number of neurons increases, we hold $m$ and $\chi$ fixed while allowing $h$ and $\lambda$ to vary with $N$. Taking the total derivative with respect to $N$ gives
\begin{equation}
0 = \frac{d}{dN} \binom{m}{\chi}  = J \frac{d}{dN} \binom{h}{\lambda} + \frac{\partial}{\partial N} \binom{m}{\chi}.
\end{equation}
Increasing the number of neurons therefore induces the following flow in the space of parameters,
\begin{equation}
\frac{d}{dN} \binom{h}{\lambda} = -J^{-1}\frac{\partial}{\partial N} \binom{m}{\chi}.
\end{equation}
In Fig.~\ref{fig_3}b, we plot the flow in parameters with respect to fractional changes in system size,
\begin{equation}
\frac{d}{d \ln N} \binom{h}{\lambda} = -NJ^{-1}\frac{\partial}{\partial N} \binom{m}{\chi}.
\end{equation}

\subsection{Mean-field approximation.} To understand the thermodynamic limit $N\rightarrow \infty$, we write the model as a function of the population activity $P(\bm{x}) \propto e^{-Nf(\mu)}$, where
\begin{equation}
f(\mu) = -h\mu - \frac{\lambda}{2} \mu^2 - \frac{1}{N}\log C\Big(N,\frac{1+\mu}{2}N\Big)
\end{equation}
is the free energy per neuron. As the system becomes large, the free energy approaches
\begin{equation}
\label{EQ:freeEnStirling}
f(\mu) = - h\mu -\frac{ \lambda }{2}\mu^2 + \frac{1+\mu}{2}\log\left(\frac{1+\mu}{2}\right) + \frac{1-\mu}{2}\log\left(\frac{1-\mu}{2}\right).
\end{equation}
For parameters $(h,\lambda)$ not on the critical line, the free energy has a unique minimum. In the limit $N\rightarrow \infty$, this minimum dominates any average over $P(\bm{x})$, yielding the mean-field theory with $m = \tanh(\lambda m + h)$ and $\chi = \frac{1-m^2}{1-\lambda(1-m^2)}$.\cite{tanaka2000information} This theory places an upper bound on the strength of correlations,\cite{di2026extended}
\begin{equation}
\chi_\text{max}(m) = \frac{m(1-m^2)}{m - \tanh^{-1}(m)(1-m^2)}.
\end{equation}
All of the recordings break this bound, some with correlations orders of magnitude stronger than allowed under mean-field theory (Fig.~\ref{fig_5}a).

\subsection{Double-well approximation.} The mean-field approximation fails to capture real neural activity because it only applies to models for which the free energy has a unique minimum; that is, non-critical models. When inferring models from neural recordings, the free energy develops two degenerate minima (Fig.~\ref{fig_5}b), the defining feature of a first-order phase transition.\cite{di2026extended} For large $N$, when taking averages over $P(\bm{x})$, we must include both minima.

For $\lambda > 1$ and $h \approx 0$, the free energy has two minima $\mu_\pm \approx \pm m_0$, where $m_0$ is the positive solution to the mean-field equation with no external field, $m_0 = \tanh(\lambda m_0)$. Only including these minima, the average activity is given by
\begin{align}
m &= \frac{\mu_+e^{-Nf(\mu_+)} + \mu_-e^{-Nf(\mu_-)}}{e^{-Nf(\mu_+)} + e^{-Nf(\mu_-)}} = \frac{m_0e^{Nhm_0} - m_0e^{-Nhm_0}}{e^{Nhm_0} + e^{-Nhm_0}} = m_0 \tanh(N h m_0).
\end{align}
Differentiating with respect to $h$ yields the correlation $\chi = N(m_0^2 - m^2)$. Finally, inverting these equations provides the solution to the inverse map from statistics to parameters in Eq.~(\ref{EQ:double_well}).

\end{methods}

\section*{Data Availability}

The data analyzed in this paper are openly available at \\ \texttt{https://github.com/dpcarcamo/Emergence-of-Criticality}.

\section*{Code Availability}

The code used to perform the analyses in this paper is openly available at \\ \texttt{https://github.com/dpcarcamo/Emergence-of-Criticality}.

%% Put the bibliography here, most people will use BiBTeX in
%% which case the environment below should be replaced with
%% the \bibliography{} command.

\newpage

%\section*{References}
\section*{\large References}
\vspace{-30pt}
\noindent\rule{\textwidth}{.5pt}

\bibliography{BibCriticality}

%% Here is the endmatter stuff: Supplementary Info, etc.
%% Use \item's to separate, default label is "Acknowledgements"
\newpage
\begin{addendum}

\item[Supplementary Information.] Supplementary text and figures accompany this paper.

\item[Acknowledgements.] We thank L.~Di Carlo, B.~Machta, Q.~Yu, and L.~Wu for enlightening discussions and comments on earlier versions of the paper. We also acknowledge support from the National Institutes of Health (NIH/NIGMS R35GM160188) and the Department of Physics, Quantitative Biology Institute, and Wu Tsai Institute at Yale University.
 
\item[Author Contributions.] D.P.C.~and C.W.L.~designed research; D.P.C.~performed research; D.P.C.~contributed
new reagents/analytic tools; D.P.C.~analyzed data; and
D.P.C.~and C.W.L.~wrote the paper.
 
\item[Competing Interests.] The author declares no competing financial interests.
 
\item[Corresponding Author.] Correspondence and requests for materials should be addressed to C.W.L. \\(christopher.lynn@yale.edu).
 
\end{addendum}

\end{document}

\myfont

%\linenumbers

\maketitle

%\begin{affiliations}
%\item Department of Physics, Quantitative Biology Institute, and Wu Tsai Institute, Yale University, New Haven, CT 06520, USA
%\item Initiative for the Theoretical Sciences, The Graduate Center, City University of New York, New York, NY 10016, USA
%\item Joseph Henry Laboratories of Physics and Lewis-Sigler Institute for Integrative Genomics, Princeton University, Princeton, NJ 08544, USA
%\item Princeton Neuroscience Institute, Princeton University, Princeton, NJ 08544, USA
%\item Center for Studies in Physics and Biology, Rockefeller University, New York, NY 10065, USA
%\item Department of Organismal Biology and Anatomy and Department of Physics, University of Chicago, Chicago, IL 60637, USA
%\end{affiliations}

%\newpage

%{\spacing{.8} \fontsize{9}{9} \tableofcontents}
\vspace{40pt}
\tableofcontents

\newpage

\section{Introduction}

In this Supplementary Information, we provide additional analyses that support and extend the results presented in the main text. In Sec.~\ref{Sec:bound}, we derive the finite-size bounds on the model statistics and identify the region of statistics space that can be realized by the Curie--Weiss model. In Sec.~\ref{Sec:Traj}, we examine how inferred model parameters evolve with increasing system size and provide additional examples of the resulting trajectories through parameter space. In Sec.~\ref{Sec:Thermo}, we characterize the thermodynamic signatures of the inferred models using several complementary perturbations and measures of criticality. Finally, in Sec.~\ref{Sec:ind}, we study how the model behaves under independent data.

\section{Bounds on statistics for finite system size}
\label{Sec:bound}

In the main text, we infer model parameters from points in statistics space. Although the parameters $h$ and $\lambda$ are unbounded, a finite system cannot realize arbitrary values of the mean activity $m$ and correlation $\chi$. For binary variables $x_i\in\{-1,1\}$, the instantaneous population activity
\begin{equation}
    \mu(\bm{x})=\frac{1}{N}\sum_{i=1}^{N}x_i
\end{equation}
satisfies $-1\leq\mu\leq1$, and the mean activity is $m=\langle\mu\rangle$. Since
\begin{equation}
    \chi=N\left(\langle\mu^2\rangle-m^2\right),
\end{equation}
the correlation is nonnegative. For a fixed mean $m$, the variance is maximized by a distribution supported at the endpoints $\mu=\pm1$, with probabilities
\begin{equation}
    P(\mu=\pm1)=\frac{1\pm m}{2}.
\end{equation}
Because $\langle\mu^2\rangle=1$ for this distribution, the maximum correlation is
\begin{align}
    \chi_{\max}
    &=N\left(\langle\mu^2\rangle-m^2\right)\\
    &=N\left(1-m^2\right).
\end{align}
Thus, the finite-size statistics satisfy
\begin{equation}
    -1\leq m\leq1,
    \qquad
    0\leq\chi\leq N\left(1-m^2\right).
    \label{EQ:statistics_bound}
\end{equation}
All points used in the fixed-statistics analyses were chosen to satisfy Eq.~(\ref{EQ:statistics_bound}) for every system size considered.

\section{Trajectories of inferred parameters with increasing system size}
\label{Sec:Traj}

The main text defines a flow field that describes how the inferred parameters $(h,\lambda)$ change with system size while the statistics $(m,\chi)$ are held fixed. Here we integrate that flow to show complete parameter trajectories. Starting from an initial condition $(m_0,\chi_0,N_0)$, we increase $N$ and follow the corresponding inferred parameters.

\begin{figure}[t]
\centering
\includegraphics[width = \textwidth]{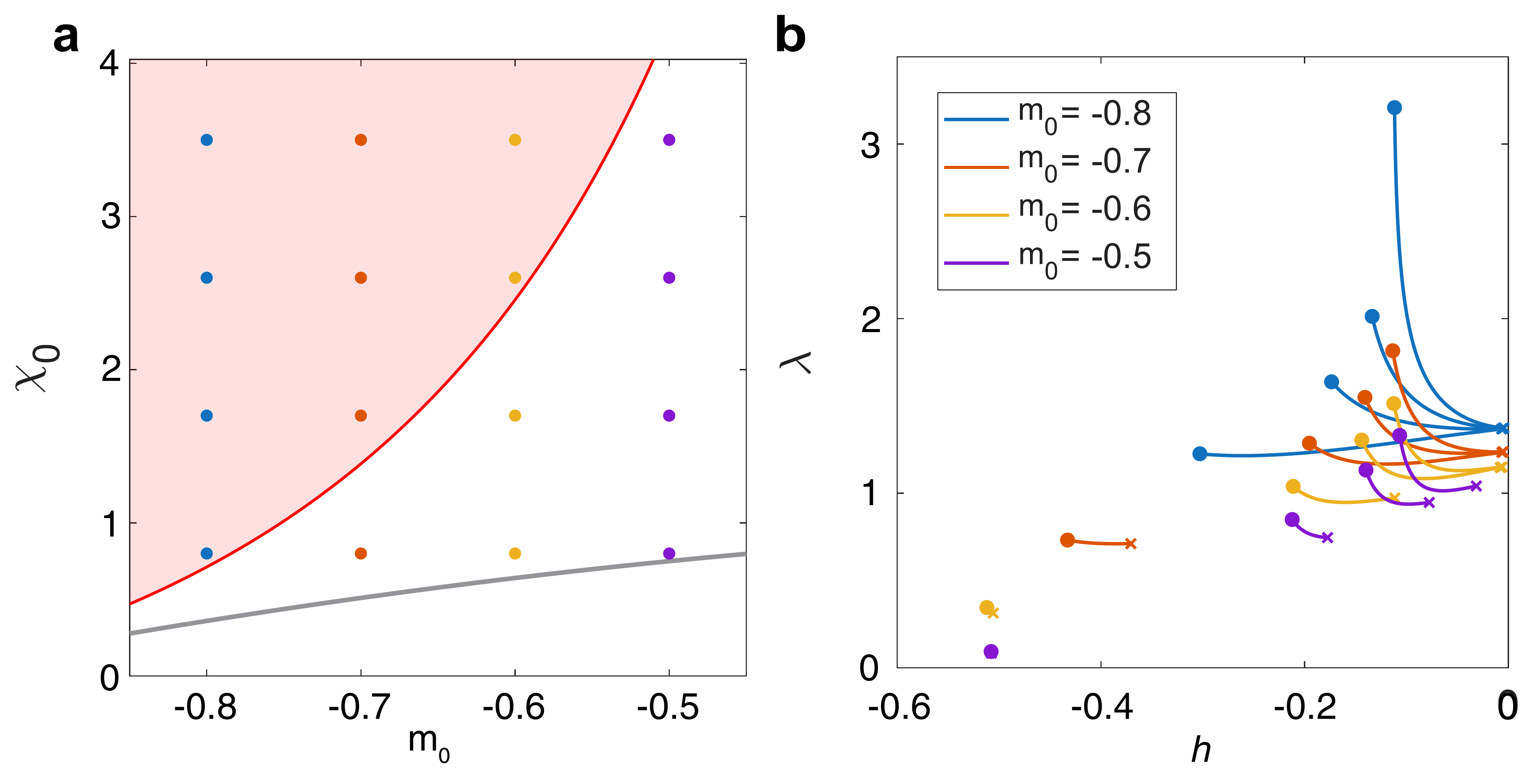} \\
\raggedright
\captionsetup{labelformat=empty}
{\spacing{1.25} \caption{\small \myfont \textbf{Fig.~S\ref{fig_traj} $|$ Trajectories of inferred parameters as system size increases at fixed statistics.} \textbf{a}, Initial conditions in statistics space. Color indicates the mean activity $m_0$. The red region marks statistics for which the mean-field approximation breaks down, as shown in Fig.5a. \textbf{b}, Parameter trajectories obtained by integrating the flow from the initial conditions in \textbf{a}, beginning at $N_0=10$. Initial conditions in the red region generate trajectories that approach the critical line, whereas points outside this region approach distinct parameter values consistent with the mean-field theory. \label{fig_traj}}}
\end{figure}

The trajectories reveal two regimes in statistics space (Fig.~S\ref{fig_traj}). Initial conditions in the red region, where the mean-field approximation discussed in the main text breaks down (Fig.5a), produce trajectories that approach the critical line $h=0$, $\lambda>1$. Outside this region, the trajectories approach distinct parameter values consistent with mean-field theory. These results complement the local flow field in Fig.3b by showing the integrated paths of inferred models as $N$ increases.

\section{Thermodynamic signatures of the inferred models}
\label{Sec:Thermo}

\begin{figure}
    \centering
    \includegraphics[width=\linewidth]{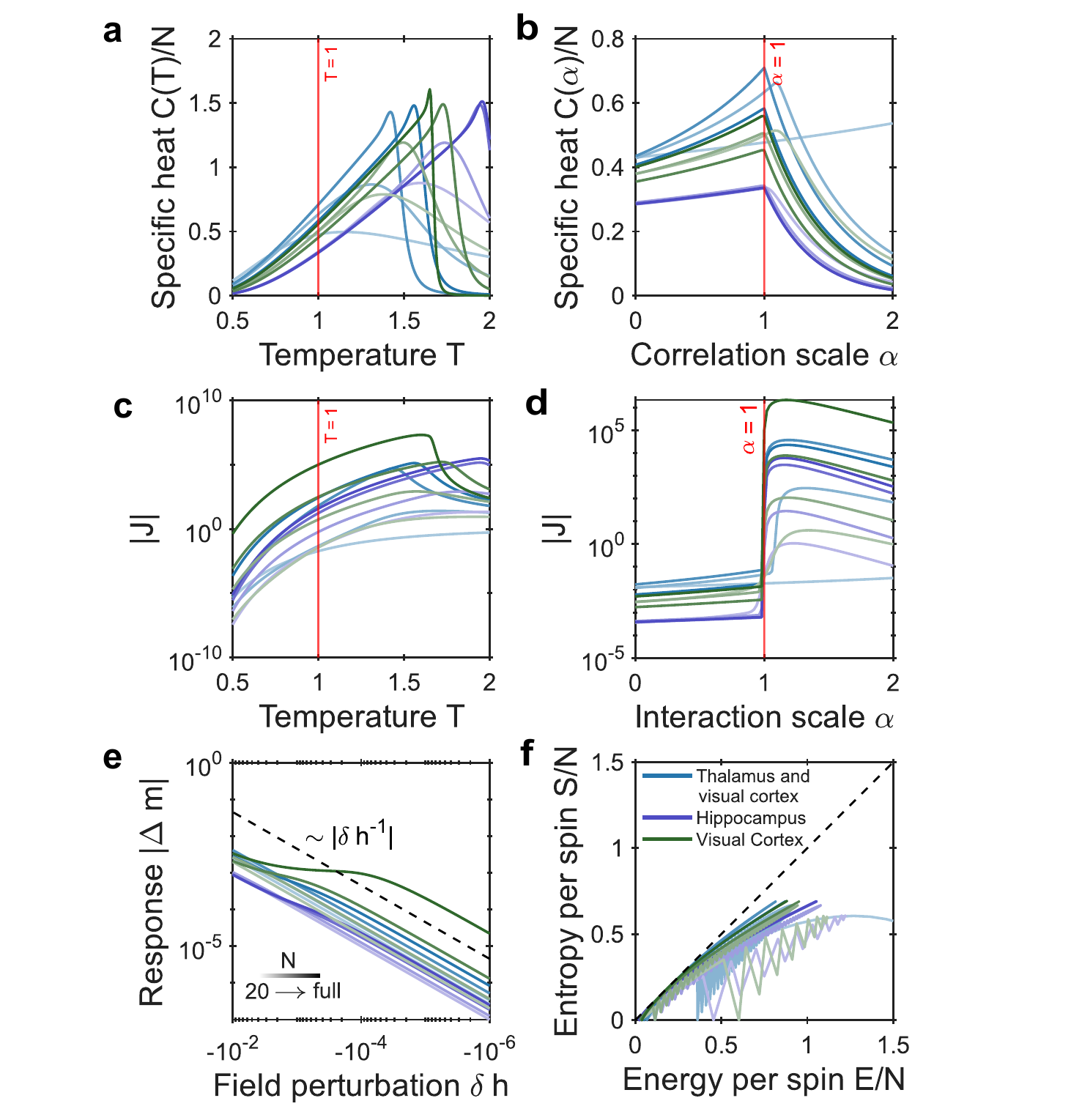}
    \captionsetup{labelformat=empty}
    {\spacing{1.25}\caption{\small \myfont \textbf{Fig.~S\ref{fig_sig} $|$ Thermodynamic signatures of the inferred Curie--Weiss models.} The three populations studied in the main text are plotted in blue, green, and purple for the thalamus and visual cortex, visual cortex, and hippocampus data, respectively. The population sizes are $N=20$, $100$, $1000$, and the full population. \textbf{a}, Specific heat, $C(T)/N$, along the fictitious-temperature path $(h/T,\lambda/T)$. The peak lies near, but not exactly at, the inferred model $T=1$. \textbf{b}, Specific heat along the fixed-mean path, with $\lambda(\alpha)=\alpha\lambda$ and $h(\alpha)$ chosen to preserve $m$. Along this path, the inferred model at $\alpha=1$ lies at the point of maximal sensitivity. \textbf{c}, Magnitude of the Jacobian determinant, $|J(T)|$, along the fictitious-temperature path. The inferred model lies close to, but slightly displaced from, the maximum. $\quad\quad\quad\quad\quad$ \label{fig_sig}}}
\end{figure}

For each neural dataset and population size, we inferred the Curie--Weiss parameters $(h,\lambda)$ from the measured mean activity $m$ and correlation $\chi$. Because these parameters define an exact finite-size model, we can perturb the inferred model and study the response of the statistics (Fig.~S\ref{fig_sig}).

We first introduce a fictitious temperature $T$ by rescaling both inferred parameters,
\begin{equation}
    h(T)=\frac{h}{T},
    \qquad
    \lambda(T)=\frac{\lambda}{T}.
\end{equation}
The original inferred model is recovered at $T=1$. A peak in the specific heat $C(T)/N$, where $C(T)$ is the heat capacity, is a signature that an inferred maximum-entropy model is poised near a second-order phase transition.\cite{tkacik_thermodynamics_2015} In our models, the maximum of $C(T)/N$ occurs at $T > 1$ (Fig.~S\ref{fig_sig}a). This reflects the fact that the inferred models lie near a first-order, rather than second-order, transition.

\addtocounter{figure}{-1}
\begin{figure}[t!]
\centering
\raggedright
\captionsetup{labelformat=empty}
{\spacing{1.25} \caption{\small \myfont \textbf{d}, Magnitude of the Jacobian determinant along the fixed-mean path, which is maximized at $\alpha=1$. \textbf{e}, Change in mean activity, $\Delta m$, produced by perturbing the inferred external field $h$ while holding $\lambda$ fixed. Solid curves show the response of the models, and the dotted curve shows the predicted scaling of the response as a function of the external field from the double-well approximation. \textbf{f}, Entropy per neuron versus energy per neuron for the inferred models. The approach to a unit-slope relation indicates that the increasing number of higher-energy states compensates for their decreasing probability. In \textbf{a--d}, red vertical lines mark the unperturbed model, corresponding to $T=1$ or $\alpha=1$.}}
\end{figure}

We next perturb the interaction strength according to
\begin{equation}
    \lambda(\alpha)=\alpha\lambda,
\end{equation}
with $\alpha=1$ corresponding to the inferred model, and choose $h(\alpha)$ so that the mean activity $m$ remains fixed. This perturbation removes the change in mean activity that accompanies the fictitious temperature. Along this fixed-mean path, the specific heat is maximized at $\alpha=1$ (Fig.~S\ref{fig_sig}b).

The same distinction between the two different parameter sweeps appears in the Jacobian $ J=\partial(m,\chi)/\partial(h,\lambda)$. Along the fictitious-temperature path, $|J(T)|$ is maximized at $T > 1$ (Fig.~S\ref{fig_sig}c). By contrast, along the fixed-mean path, $|J(\alpha)|$ peaks at $\alpha=1$ (Fig.~S\ref{fig_sig}d). This result directly supports the geometric interpretation developed in the main text: when perturbations are constrained to preserve the measured mean activity, the inferred model lies at a point where small parameter changes produce especially large changes in the remaining model statistics.

Fig.~S\ref{fig_sig}e tests the response of the inferred models to perturbations of the external field. For each inferred model, we vary $h$ around its inferred value while holding $\lambda$ fixed and calculate the resulting change in mean activity, $\Delta m$. The solid curves show the responses calculated from models with different population sizes. The dotted line shows the predicted scaling from the double-well approximation derived in the main text. Their agreement shows that the strong response of the mean activity to small field perturbations is captured by the double-well approximation to the free energy near the critical line.

Finally, we compare entropy and energy across the inferred models (Fig.~S\ref{fig_sig}f). In an equilibrium representation, the number of states at energy $E$ grows as $\exp[S(E)]$, while their Boltzmann weight at $T=1$ decreases as $\exp(-E)$. The competition between these terms selects a narrow range of typical energies. If instead $S(E)$ approaches a line with unit slope, the growth in the number of states compensates for the decrease in their probability over a broad energy range. This entropy--energy balance was identified as a signature of criticality in neural maximum-entropy models.\cite{tkacik_thermodynamics_2015} The inferred Curie--Weiss models approach this relation as population size increases, providing an additional indication that they lie close to a critical thermodynamic structure. Together, Fig.~S\ref{fig_sig}a-f show that the response of the system depends on the direction in which the inferred model is perturbed, consistent with the proximity to a first-order phase transion observed in the main text.

\section{Independent data}
\label{Sec:ind}

\begin{figure}[t]
\centering
\includegraphics[width = \textwidth]{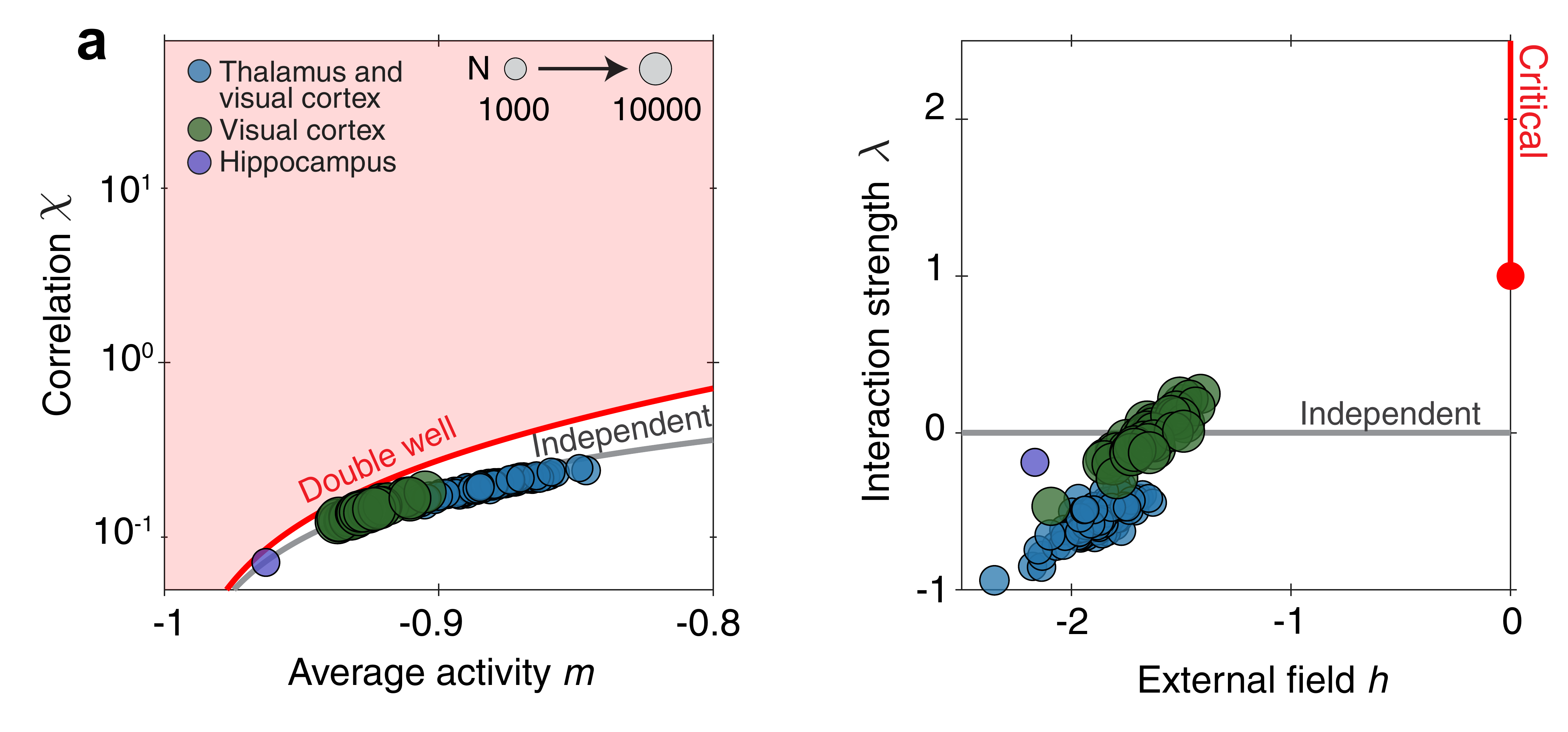} \\
\raggedright
\captionsetup{labelformat=empty}
{\spacing{1.25} \caption{\small \myfont \textbf{Fig.~S\ref{fig_shuffle} $|$ Shuffled data.} The time series of each neuron was independently shuffled, preserving the mean activity of each neuron while removing correlations between neurons in expectation. Left, the measured $m$ and $\chi$ for the shuffled data. The gray line shows the statistics of the independent Curie--Weiss model ($\lambda=0$), $\chi=1-m^2$. Shuffling moves the measured statistics toward the region expected for independent activity. Because the shuffled time series have finite length, residual correlations due to finite sampling can place some points above the independent curve. Right, the corresponding Curie--Weiss parameters $h$ and $\lambda$ inferred from the shuffled statistics. Shuffling moves the inferred models toward the non-interacting model, $\lambda=0$.\label{fig_shuffle}}}
\end{figure}

To determine how correlations in the neural activity affect the inferred models, we shuffle the time series of each neuron independently. This procedure preserves the mean activity of each neuron while removing correlations between neurons in expectation. We then calculate $m$ and $\chi$ from the shuffled time series and infer the corresponding Curie-Weiss parameters $h$ and $\lambda$ in the same way as for the original data. Shuffling moves the measured statistics decreases the total correlation $\chi$ and correspondingly moves the inferred models toward $\lambda=0$ (Fig.~S\ref{fig_shuffle}).

Since the shuffled time series have finite length, however, the measured pairwise correlations do not vanish exactly. These finite-sampling fluctuations can cause some shuffled datasets to lie above the independent Curie--Weiss curve, $\chi=1-m^2$. To remove this finite-sampling effect, we therefore considered a second control in which we assumed exact independence and calculated $\chi$ directly from the measured mean activity of each neuron.

\begin{figure}[t]
\centering
\includegraphics[width = \textwidth]{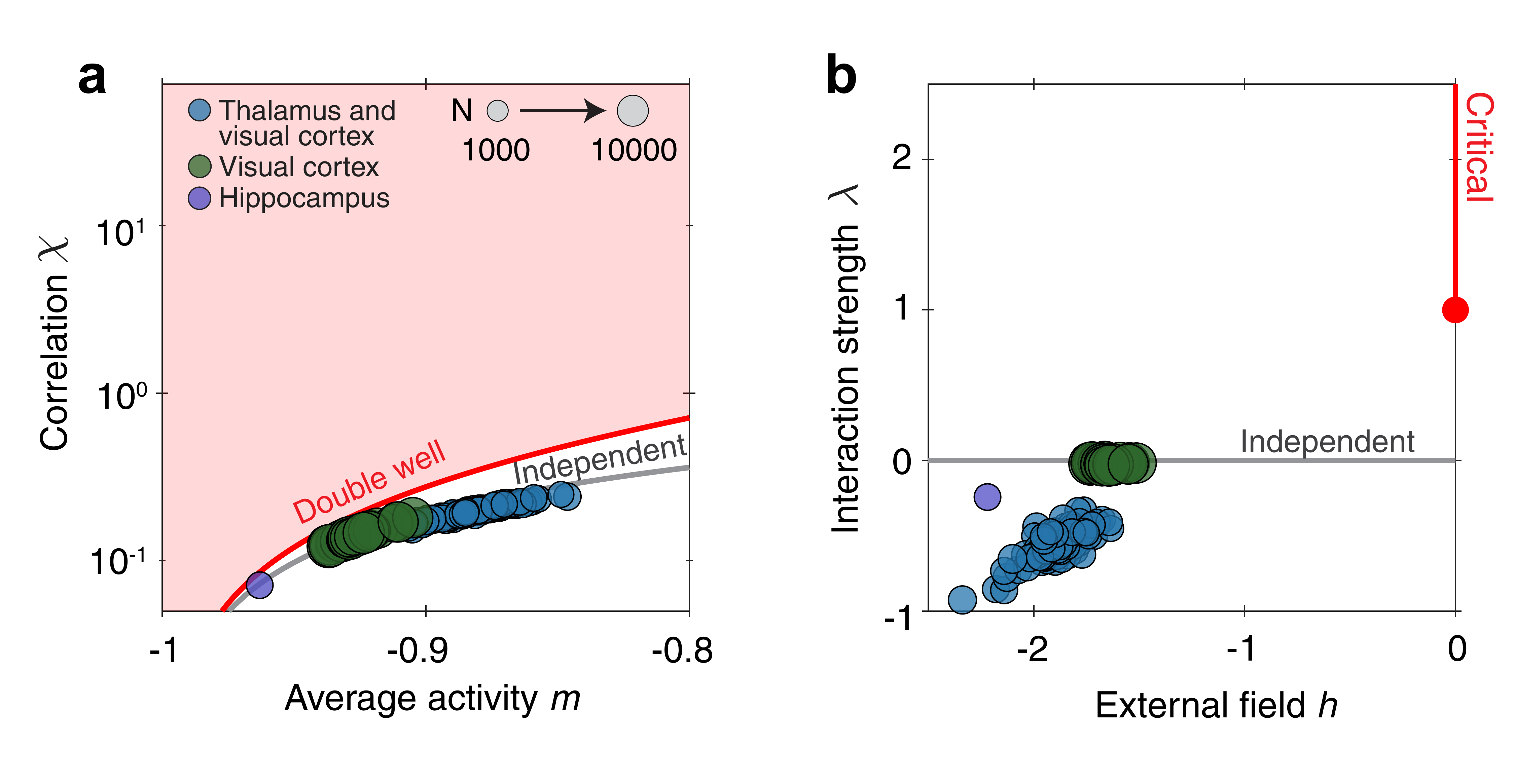} \\
\raggedright
\captionsetup{labelformat=empty}
{\spacing{1.25} \caption{\small \myfont \textbf{Fig.~S\ref{fig_inddata} $|$ Independent data.} The statistics expected under exact independence were calculated directly from the measured mean activity of each neuron, with all off-diagonal covariances set to zero. Left, the resulting $m$ and $\chi$. The gray line shows the independent Curie--Weiss model ($\lambda=0$), $\chi=1-m^2$. Heterogeneity in the mean activity of individual neurons reduces the correlation to $\chi=1-m^2-\Delta_x^2$, where $\Delta_x^2=N^{-1}\sum_i(\langle x_i\rangle-m)^2$. The statistics therefore lie at or below the independent Curie--Weiss curve when finite-sampling correlations are removed. Right, the corresponding $h$ and $\lambda$ obtained by mapping these statistics onto the Curie--Weiss model.\label{fig_inddata}}}
\end{figure}

For independent neurons, all off-diagonal covariances vanish, giving
\begin{equation}
    \chi
    = \frac{1}{N}\sum_i
    \left(\langle x_i^2\rangle-\langle x_i\rangle^2\right)
    = 1-\frac{1}{N}\sum_i\langle x_i\rangle^2,
\end{equation}
where we have used $x_i^2=1$. Defining the average squared deviation of the individual neuron means from the population mean as
\begin{equation}
    \Delta_x^2
    = \frac{1}{N}\sum_i
    \left(\langle x_i\rangle-m\right)^2,
\end{equation}
and using $m=N^{-1}\sum_i\langle x_i\rangle$, we have
\begin{equation}
    \frac{1}{N}\sum_i\langle x_i\rangle^2
    =m^2+\Delta_x^2.
\end{equation}
The correlation of an independent population is therefore
\begin{equation}
    \chi=1-m^2-\Delta_x^2.
\end{equation}
Since $\Delta_x^2\geq0$, an independent population with heterogeneous mean activities $\langle x_i\rangle$ must satisfy
\begin{equation}
    \chi\leq1-m^2,
\end{equation}
with equality only when all neurons have the same mean activity. When exact independence is imposed, all of the datasets therefore lie at or below the independent Curie-Weiss curve (Fig.~S\ref{fig_inddata}). When inferring Curie-Weiss models, this yields non-positive interactions $\lambda \le 0$.

\newpage

\section*{References}
\addcontentsline{toc}{section}{References}

\vspace{24pt}

\bibliography{BibCriticality}

\newpage
\clearpage